\documentclass[12pt]{iopart}
\usepackage{url}
\usepackage{hyperref}
\usepackage{graphicx}
\usepackage{caption}
\usepackage{amssymb} 

\usepackage{xcolor}
\usepackage{ulem}
\usepackage{cite}

\begin{document}

\title{Monte Carlo event generation of neutrino-electron scattering} 

\author{Dmitry Zhuridov$^1$
	, Jan T. Sobczyk$^1$
	, Cezary Juszczak$^1$
	and Kajetan Niewczas$^{1,2}$
}

\address{$^1$Institute  of  Theoretical  Physics, University of Wroc{\l}aw, Plac  Maxa  Borna  9,  50-204,  Wroc{\l}aw,  Poland}
\address{$^2$Department of Physics and Astronomy, Ghent University, Proeftuinstraat 86, B-9000 Gent, Belgium}
\eads{\mailto{dmitry.zhuridov@uwr.edu.pl}, \mailto{jan.sobczyk@uwr.edu.pl}}
%
\vspace{10pt}
\begin{indented}
\item[]22.12.2020
\end{indented}

\begin{abstract}
We describe an extension of the NuWro Monte Carlo neutrino event generator with the neutrino-electron scattering processes. This new dynamical channel includes the charged current and neutral current interactions, together with their interference, for $\nu_\ell\, \, e$ and $\bar\nu_\ell\, \, e$ ($\ell=e,\,\mu,\,\tau$) scatterings, resulting in ten possible final states. We illustrate the performance of the new functionality on few physical examples, including an estimation of the background in the T2K $\nu_\mu\rightarrow \nu_e$ oscillation experiment. We show that the background events arising from the neutrino-electron interactions occupy mostly a distinct region of the phase space and can be easily separated.
\end{abstract}

\vspace{2pc}
\noindent{\it Keywords}: neutrino interactions, Monte Carlo simulations, neutrino oscillations,  neutrino electron scattering, antineutrino electron scattering, neutrino flavour
%
%
\maketitle
%

\section{Introduction}

Neutrino flavour oscillation experiments provide a unique opportunity to investigate basic assumptions of the Standard Model (SM) of elementary particles and its extensions, including measurements of the yet unknown CP-violating phase in the leptonic sector~\cite{Abe:2019vii}. These experiments strongly rely on precision in describing neutrino interactions with matter. Although the majority of interests lie in neutrino-nucleus reactions, experimental signals also contain a small contribution from rare neutrino-electron interactions. The success of future high statistics long-baseline experiments such as DUNE~\cite{Abi:2020evt} and Hyper-Kamiokande~\cite{Abe:2015zbg} 
depends on a substantial reduction of all sources of systematic uncertainties. Thus, the 
inclusion of neutrino-electron scattering events in the oscillation analyses may become relevant. Apart from the long-baseline experiments, which use accelerator neutrino beams in the few-GeV energy region, neutrino-electron interactions play a vital role in studies of solar (e.g., the Super-Kamionade~\cite{Fukuda:2001nj}, SNO~\cite{BOGER2000172-short}, and Borexino~\cite{Alimonti:2008gc} experiments) and supernovae neutrinos~\cite{Bionta:1987qt} with energies in the $\sim 1-50$~MeV range.

Monte Carlo (MC) neutrino event generators~\cite{Tanabashi:2018oca} are indispensable tools in the neutrino oscillation analyses, providing theoretical predictions of neutrino interactions in particular experimental environments, accounting for the specific neutrino fluxes and detector complexity. There are several independently developed codes, such as NEUT~\cite{Hayato:2009zz} and GENIE~\cite{Andreopoulos:2009rq}, which are used for the actual oscillation analyses and optimized for the energy range of accelerator neutrinos. Other MC generators or simulation tools, such as NuWro~\cite{NuWro} and GiBUU~\cite{GiBUUrev}, developed by theorists, have been used in numerous comparisons and neutrino cross section studies. Monte Carlo neutrino event generators, serving as a bridge between theoretical models and experimental measurements, have to be further improved to facilitate the progress in neutrino physics~\cite{Alvarez-Ruso:2017oui}.

In this work, we describe an implementation of the neutrino-electron interaction dynamics in the NuWro generator, extending the range of its possible applications and increasing the precision of the background estimation in neutrino-nucleus scattering. MC implementation of such a channel has been recently discussed in Ref.~\cite{Grichine:2019vcc}, aiming to add this interaction mode into the GEANT4 package. This software predominantly conducts detector simulations, where neutrino interactions have no practical importance. The neutrino event generators GENIE and NEUT also have this interaction channel implemented, yet the actual oscillation studies usually neglect it. Neutrino-electron scattering, being a purely leptonic process, can be calculated directly from the SM. Its interaction cross section is precisely known, offering a clean experimental signature with a uniquely small cross section uncertainty. Thus, neutrino interactions with atomic electrons, despite their rarity, are used in long-baseline experiments for constraining neutrino fluxes~\cite{Park:2015eqa, Valencia:2019mkf}, and it is essential to have neutrino-electron scattering mechanism implemented in the used Monte Carlo simulation tools.

Having validated the new NuWro functionality with a series of necessary tests, to demonstrate the advantages of having the new interaction channel in a Monte Carlo neutrino event generator, we performed a study in the context of the T2K~\cite{Abe:2011ks} long-baseline oscillation experiment. We calculated the contribution from neutrino-electron events in the overall sample of electron events registered in the Super-Kamiokande detector. Then, we showed that in the future Hyper-Kamiokande experiment, among hundreds of events with an electron in the final state, as used in the future investigations of the $\nu_\mu\rightarrow \nu_e$ appearance signal, there should exist events coming from the $\nu +e\rightarrow\nu +  e$ processes, but with distinct kinematical characteristics that allow for their clear identification and rejection. 

Our paper is organized as follows. In Section~\ref{sec:Analytics}, the analytical expressions for neutrino scattering off electrons are summarized. Then, in Section~\ref{sec:NuWro_update}, we describe shortly the construction of the NuWro Monte Carlo generator and explain technical details of the new functionality. In Section~\ref{sec:Applications}, we present a few validation tests of the discussed implementation and exemplify its application in the evaluation of the leptonic background in the T2K oscillation experiment. We conclude in Section~\ref{sec:Conclusion} with our final remarks.

\section{Neutrino scattering off charged leptons}
\label{sec:Analytics}

\subsection{Lagrangian}

\begin{figure}
	\centering
	\includegraphics[width=0.35\textwidth]{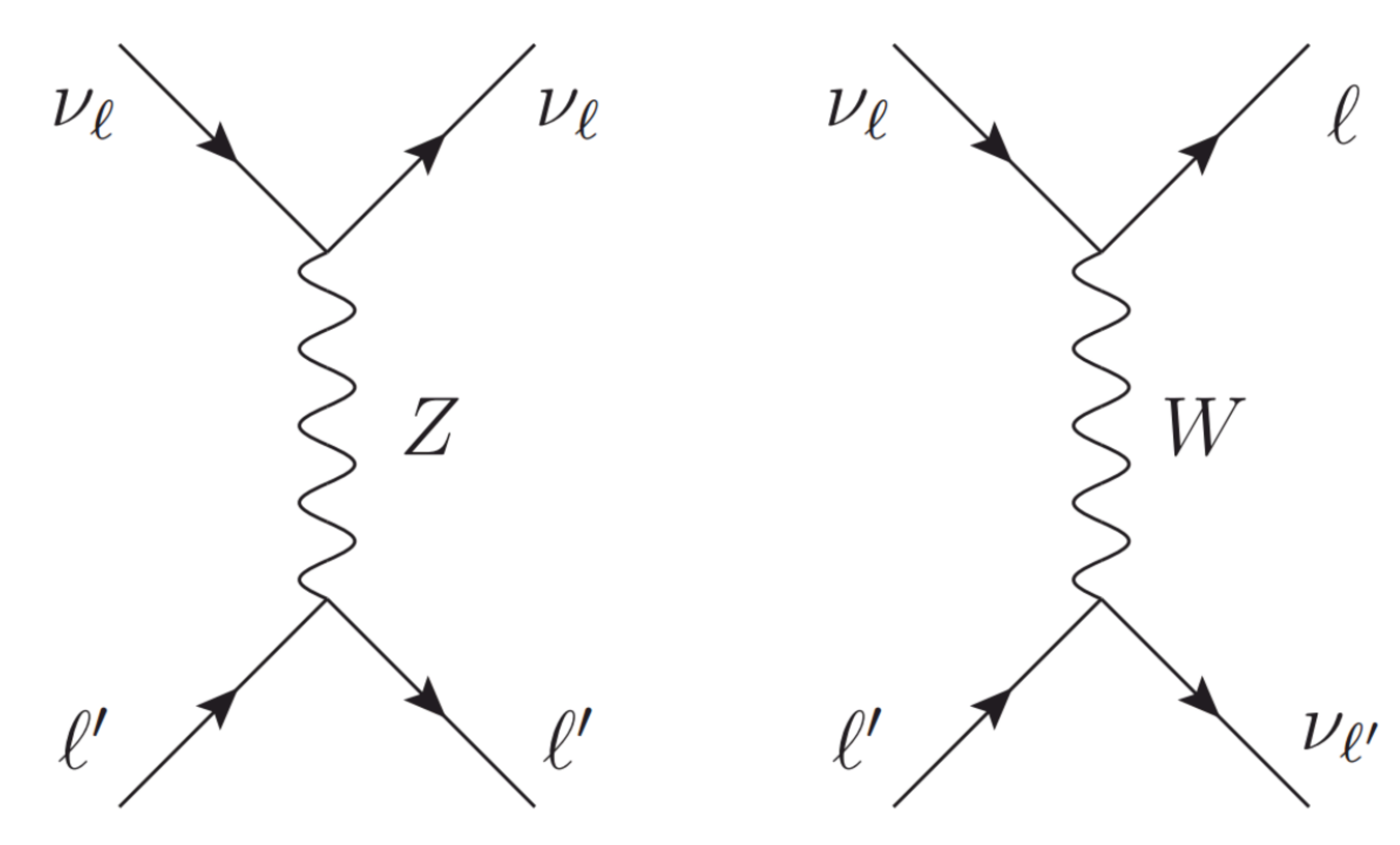} \hspace{3mm} 
	\includegraphics[width=0.29\textwidth]{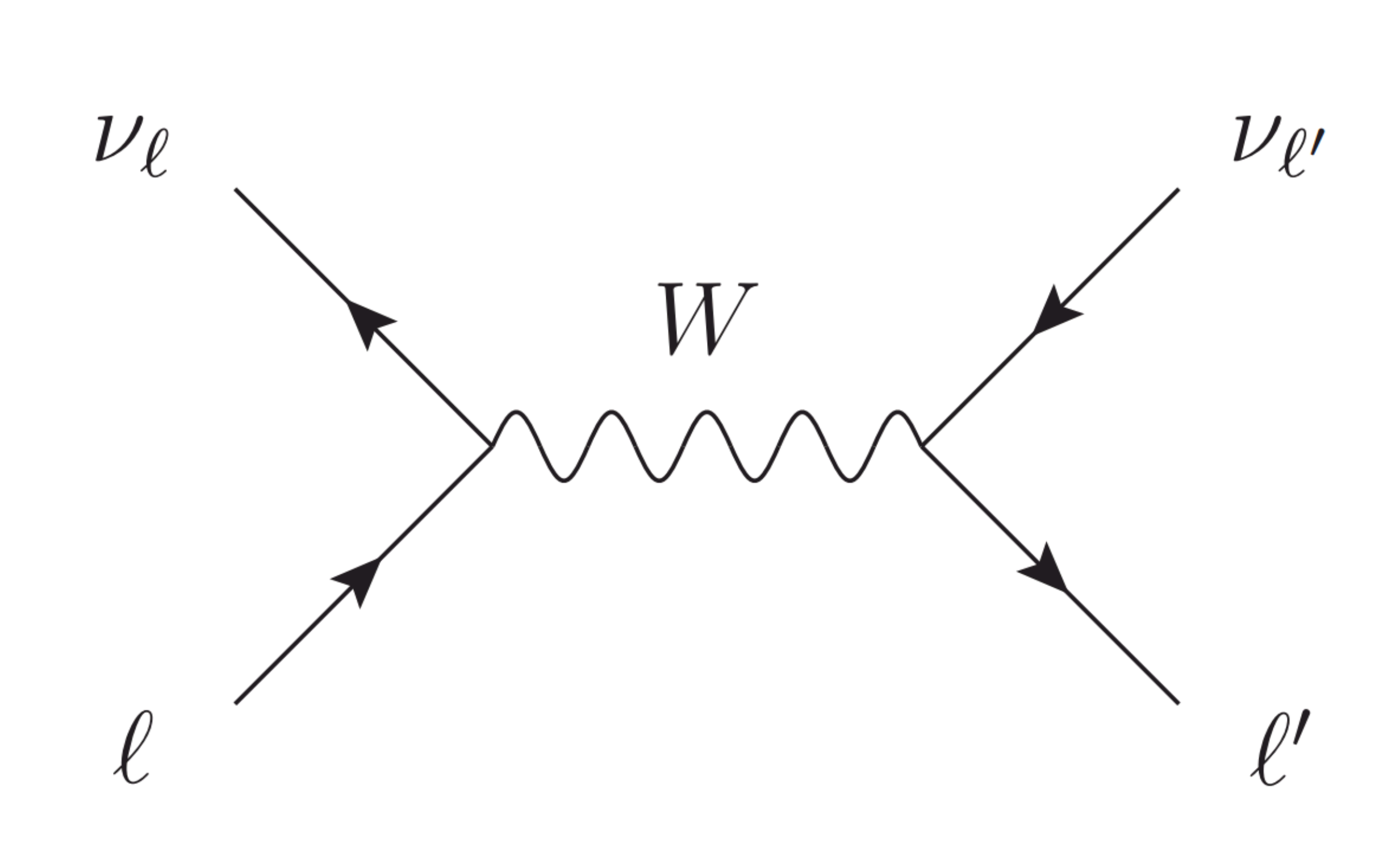}
	\vspace{-3mm}
	\caption{~Feynman diagrams with leading contributions to neutrino-charged lepton scattering.}
		\label{Fig:diagrams}
\end{figure}

In the Standard Model, neutrinos scatter off charged leptons via $W^\pm$ and $Z^0$ boson exchanges, which is represented by the Feynman diagrams shown in Fig.~\ref{Fig:diagrams}. Neglecting the intermediate bosons four-momenta, much smaller compared to their masses, the tree level low-energy effective Lagrangian for this interaction~\cite{Tomalak:2019ibg,Fermi:1934hr} can be written as
\begin{eqnarray}
	\mathcal{L}_{\rm eff} \ = 
	-c_0\left[\sum\limits_{\ell,\ell^\prime} N^\mu_{\ell\ell} \, \bar\ell^\prime\gamma_\mu(c_L^{\ell\ell^\prime}P_L+c_RP_R)\ell^\prime 
	+ \sum\limits_{\ell\neq\ell^\prime} N^\mu_{\ell^\prime\ell} \,
	\bar\ell\gamma_\mu P_L \ell^\prime \right],
\end{eqnarray}
where $N^\mu_{\ell^\prime\ell} = \bar\nu_{\ell^\prime}\gamma^\mu P_L\nu_\ell$ is the neutrino current, $\ell=e, \mu, \tau$ denotes the charged lepton, and $\nu_\ell$ is a neutrino of the respective flavour. $P_L$ and $P_R$ are the left and right chirality projectors. The numerical coefficients are
\begin{eqnarray}
	c_0=2\sqrt{2}{\rm G}_{\rm F}, \qquad c_L^{\ell\ell^\prime} = \sin^2\theta_W-\frac{1}{2}+\delta_{\ell\ell^\prime}, \qquad c_R = \sin^2\theta_W,
\end{eqnarray}
where $\delta_{\ell\ell^\prime}$ is the Kronecker symbol, $\theta_W$ is the Weinberg angle, and
\begin{eqnarray}
    {\rm G}_{\rm F}=\frac{g^2}{4\sqrt{2}M_W^2}
\end{eqnarray}
is the Fermi constant relating the weak coupling constant $g$ and $W$-boson mass $M_W$.  In this paper, we do not discuss non-standard neutrino interactions~\cite{Healey:2013vka}.

In what follows, we focus on the neutrino-electron scattering processes. There are six channels corresponding to all possible incoming neutrino and antineutrino flavours, with a total of 10 subchannels differing by the set of final leptons: 
\begin{eqnarray}\label{eq:channels}
	\nu_e~e \to \nu_e~e, \quad \nu_\mu~e \to \{\nu_\mu~e,~\nu_e~\mu\}, \quad \nu_\tau~e \to \{\nu_\tau~e,~\nu_e~\tau\}, \nonumber\\
	\bar\nu_e~e \to \{\bar\nu_e~e,~\bar\nu_\mu~\mu, ~\bar\nu_\tau~\tau\}, \quad \bar\nu_\mu~e \to \bar\nu_\mu~e, \quad \bar\nu_\tau~e \to \bar\nu_\tau~e.
\end{eqnarray}	
The six processes with final states identical to the initial ones are called `elastic' in the literature. 

In these considerations, we safely neglect neutrino masses.
The atomic binding energy of electrons $10\,{\rm eV}\lesssim E_e^{\rm bind}\lesssim10\,{\rm keV}$ is much smaller than the electron mass $m_e$, and to good approximation, one can take the target electrons to be at rest. We also assume that the polarization of the target and outgoing electrons is not measured.

For the initial particles four-momenta: $p_\nu = (E_\nu,\vec p_\nu)$, $|\vec p_\nu|=E_\nu$ and $p_e = (m_e,\vec 0)$, the invariant mass squared is
\begin{eqnarray}
	s\equiv(p_\nu+p_e)^2 = 2E_\nu m_e + m_e^2,
\end{eqnarray}
and production of the charged lepton $\ell$ requires neutrino beam energy above the threshold values of
\begin{eqnarray}\label{eq:thresholds}
	E_\nu^{\ell ,thr} = \frac{m_\ell^2-m_e^2}{2m_e}.
\end{eqnarray}
In particular, for muon and tau lepton production, the respective thresholds are $E_\nu^{\mu ,thr}\simeq10.9$~GeV and $E_\nu^{\tau ,thr}\simeq3.09$~TeV. There is no neutrino energy threshold for the elastic scattering off an electron. Similarly, at the energy of
\begin{eqnarray}\label{Glashow_resonance}
	E_\nu^W = \frac{M_W^2-m_e^2}{2m_e}\approx \frac{M_W^2}{2m_e}\approx6.3~{\rm PeV},
\end{eqnarray}
resonant exchange of the $W$-boson can occur in electron antineutrino scattering off an atomic electron. For this characteristic Glashow resonance~\cite{Glashow:1960zz,Garcia:2020jwr} energy, the $\bar\nu_e$-$e$ interaction dominates over the $\bar\nu_e$-nucleus one. We plan to add this PeV-energy effect in a future NuWro update.

\subsection{Scattering cross section}

The leading order (LO) differential cross section of the neutrino-electron scattering~\cite{Tomalak:2019ibg,tHooft:1971ucy,Sarantakos:1982bp} can be written as
\begin{equation}\label{eq:diff_cross_sec1}
\frac{d}{dz}\sigma_{\rm LO}^{\nu_\ell[\bar\nu_\ell]e\to\nu_\ell[\bar\nu_\ell]e} = \frac{E_\nu m_e}{4\pi} \left\{ (c_L^{\ell e})^2I_{L[R]} + c_R^2I_{R[L]} + c_L^{\ell e} c_R I^L_R \right\} \label{eq:diff_cross_sec1a}
\end{equation}
for the elastic processes, and
\begin{equation}\left.\frac{d}{dz}\sigma_{\rm LO}^{\nu_\ell e\to\nu_{e} \ell}\right|_{\ell\neq e} = c_0^2 \frac{E_\nu m_e}{4\pi} I_L, 
	\qquad \left.\frac{d}{dz}\sigma_{\rm LO}^{\bar\nu_e e\to\bar\nu_{\ell} \ell}\right|_{\ell\neq e} = c_0^2\frac{E_\nu m_e}{4\pi} I_R \label{eq:diff_cross_sec2}
\end{equation}
for the remaining reactions listed in Eq.~(\ref{eq:channels}). 
In the formulas above,
the final to initial neutrino energy ratio denoted as $z\equiv E_\nu^\prime/E_\nu$ varies in the range of
\begin{eqnarray}
	z_{\rm min}\equiv\frac{m_e}{m_e+2E_\nu} - \frac{\Delta M^2}{2E_\nu(m_e+2E_\nu)} 
	\leq z
	\leq 1 - \frac{\Delta M^2}{2E_\nu m_e}\equiv z_{\rm max},\label{phasespace}
\end{eqnarray}
where $\Delta M^2 = m_\ell^2-m_e^2$ is the charged lepton mass splitting with the final charged lepton mass of $m_\ell = m_e,\,m_\mu,\,m_\tau$.
The kinematical factors in Eqs.~(\ref{eq:diff_cross_sec1}), (\ref{eq:diff_cross_sec2}) take the form of
\begin{eqnarray}
	I_L(E_\nu) &=& 1 - \frac{\Delta M^2}{2m_eE_\nu}, \\
	I_R(E_\nu,z) &=& z^2\left(1 + \frac{\Delta M^2}{2m_eE_\nu z}\right), \\
	I^L_R(E_\nu,z) &=& - \frac{m_\ell}{E_\nu}\left(1 - z - \frac{\Delta M^2}{2m_eE_\nu}\right).
\end{eqnarray}

We write the total LO cross sections as
\begin{eqnarray}
&&\sigma_{\rm LO}^{\nu_\ell[\bar\nu_\ell]e\to\nu_\ell[\bar\nu_\ell]e} = \frac{E_\nu m_e}{4\pi} \left\{ (c_L^{\ell e})^2J_{L[R]} + c_R^2J_{R[L]} + c_L^{\ell e} c_R J^L_R \right\}, \label{eq:sigma1}\\
&&\left.\sigma_{\rm LO}^{\nu_\ell e \to\nu_{e} \ell}\right|_{\ell\neq e} = c_0^2\frac{E_\nu m_e}{4\pi} J_L, 
	\qquad \left.\sigma_{\rm LO}^{\bar\nu_e e\to\bar\nu_{\ell} \ell}\right|_{e\neq\ell} = c_0^2\frac{E_\nu m_e}{4\pi} J_R,
	\label{eq:sigma2}
\end{eqnarray}
with the dimensionless factors
\begin{eqnarray}\label{eq:J_factor}
    J(E_\nu) = \int\limits_{z_{\rm min}}^{z_{\rm max}} I(E_\nu,z) \,dz
\end{eqnarray}    
that we show explicitly in Table~\ref{Tab:1}, 
defining
\begin{eqnarray}
	F_1 = 2+r, \qquad F_2 = (2 - rR)^2, \qquad F_3 = (2+3\,r)rR,
\end{eqnarray}
and
\begin{eqnarray}
    R = \frac{\Delta M^2}{m_e^2}, \qquad  r=\frac{m_e}{E_\nu}.
\end{eqnarray}

\begin{table}
\caption{\label{Tab:1}The kinematical factors $J$ of Eqs.~(\ref{eq:sigma1})-(\ref{eq:J_factor}), in the generic form and in the limit of large incoming neutrino energy (right column).}
	\begin{indented}
	\lineup
	\item[]\begin{tabular}{@{}lccc}
			\br
			\qquad & \quad $m_\ell = m_\mu, m_\tau$ \quad & \qquad $m_\ell=m_e$ \qquad & \quad  $m_\ell = m_e, m_\mu, m_\tau$; $E_\nu\gg m_e$ \quad \\
			\mr
			$J_L$ & $\frac{\displaystyle F_2}{\displaystyle 2 F_1}$ & $\frac{\displaystyle 2}{\displaystyle F_1}$ & $1$ \\
			$J_R$ & $\frac{\displaystyle F_2}{\displaystyle 12}\left(1 - \frac{\displaystyle r^3}{\displaystyle F_1^3} + \frac{\displaystyle F_3}{\displaystyle  F_1^3}\right)$ & $\frac{\displaystyle 1}{\displaystyle 3}\left(1 - \frac{\displaystyle r^3}{\displaystyle  F_1^3}\right)$ & $\frac{\displaystyle 1}{\displaystyle 3}$ \\
			$J^L_R$ & $-\frac{\displaystyle r}{\displaystyle 2} \frac{\displaystyle F_2}{\displaystyle F_1^2} \frac{\displaystyle m_\ell}{\displaystyle m_e}$ & $-\frac{\displaystyle 2r}{\displaystyle F_1^2}$ & $-\frac{\displaystyle 1}{\displaystyle 2}\frac{\displaystyle m_\ell}{\displaystyle E_\nu}$ \\
			\br
	\end{tabular}
	\end{indented}
\end{table}

In the MC implementation considered in this paper, we do not include radiative corrections discussed in, e.g., Refs.~\cite{Tomalak:2019ibg,tHooft:1971ucy,Sarantakos:1982bp,Lee:1964jq,Marciano:2003eq}.
For the neutrino energies up to $\sim50$~GeV, the overall $\cal{O}(\alpha)$ corrections are of the order of 2-3\%, making the total cross section slightly smaller. At even higher energies, these corrections are not large, rising logarithmically with neutrino energy. More significant is the impact of radiative corrections on the spectrum of electron energies, which soften due to emission of bremsstrahlung photons. In a  MC implementation, one should explicitly include such photons while generating final states and it is a challenging task that requires a dedicated study. For most purposes, one can neglect the effects of radiative corrections in neutrino-electron scattering and use the NuWro implementation for energies up to $\sim 100$~GeV.

\section{NuWro update}\label{sec:NuWro_update}

\subsection{NuWro interaction modes}
The neutrino Monte Carlo generator NuWro~\cite{NuWro, NuWroFSI} has been developed at the Wroc{\l}aw University by a theory group since 2005. Currently, NuWro covers the neutrino energy range from $\sim 100$~MeV to $\sim 100$~GeV. It includes the following neutrino-nucleon interaction modes which can be individually switched on and off:
\begin{itemize}
	\item charged current quasi-elastic (CCQE) and neutral current elastic scattering
	\begin{eqnarray}
	\nu_\ell+n\rightarrow \ell^-+p, \quad \bar\nu_\ell+p\rightarrow \ell^++n, \ \ \rm{and} \ \ \nu_\ell+N\rightarrow \nu_\ell+N,
	\end{eqnarray}
	where $N$ is either proton ($p$) or neutron ($n$);
	\item CC/NC  single-pion production, most importantly through the excitation of the $\Delta(1232)$ resonance;
	\item CC/NC deep-inelastic scattering, defined by the condition $W_{\rm{inv}}>1.6$~GeV, where $W_{\rm{inv}}$ is the invariant hadronic mass.
\end{itemize}
For nuclear target reactions, NuWro offers many possibilities to describe the initial state bound nucleon: local and global Fermi gas, spectral functions, effective density and momentum dependent nuclear potential. Nuclear targets bring in also two new possibilities for the interaction modes:	
	\begin{itemize}
	\item CC/NC coherent pion production;
	\item CC/NC meson exchange current process.
\end{itemize}

NuWro contains a homegrown intranuclear cascade model for the final state interactions (FSI) of outgoing hadrons~\cite{Niewczas:2019fro}. In modelling purely leptonic processes, FSI is not needed.

\subsection{Neutrino-electron scattering}

We have implemented the new dynamics within the existing NuWro framework, using the differential cross sections given in Eqs.~(\ref{eq:diff_cross_sec1})-(\ref{eq:diff_cross_sec2}).  

For each neutrino-electron scattering channel with a known value of the incoming neutrino energy, at every run a point $z$ from the phase space defined in Eq.~(\ref{phasespace}) is selected at random with a uniform distribution. With this value of the variable $z$, we calculate the event `weight', for each neutrino-electron scattering channel, as the differential cross section evaluated using the formulas~(\ref{eq:diff_cross_sec1})-(\ref{eq:diff_cross_sec2}) multiplied by the Monte Carlo phase space volume $z_{max}-z_{min}$. We use the weights for two purposes. Firstly, their average value converges to the process cross section and is reported in the output file. Secondly, a sample of events of required quantity is generated through the accept-reject algorithm.

The value of $z$ contains sufficient information to determine a neutrino-electron scattering event. The event kinematics is first generated with the known beam direction and an arbitrary choice of the interaction plane. As no information about electron polarization is explored, there is no constraint on the interaction plane, and in each event, it is selected as a rotation around the neutrino beam direction by a random angle $\phi$.

Assuming that the neutrino flux is oriented along the axis ${\cal Z}$, after the rotation, the energy-momentum four-vectors of the final state neutrino and charged lepton are given as
\begin{eqnarray}
\label{z1}
    p_\nu^\prime &=& (E_\nu z,\, E_\nu  z\sin\theta_\nu\cos\phi,\, E_\nu z\sin\theta_\nu\sin\phi,\, E_\nu z\cos\theta_\nu), \\
\label{z2}
    p_\ell^\prime   &=&(E_\ell^\prime,\, -E_\nu z\sin\theta_\nu\cos\phi,\, -E_\nu z\sin\theta_\nu\sin\phi,\, E_\nu(1-z\cos\theta_\nu)),
\end{eqnarray}
where
\begin{eqnarray}
    E_\ell^\prime = E_\nu(1 - z) + m_e
\end{eqnarray}
is the final charged lepton energy, and
\begin{eqnarray}
    \cos\theta_\nu = \frac{2E_\nu^2 z + 2m_e E_\nu(z-1) + \Delta M^2
    }{2E_\nu^2 z}
\end{eqnarray}
is the cosine of the angle between the incoming and outgoing neutrino momenta.

From the channel list in Eq.~(\ref{eq:channels}), we see that for the $\nu_\mu$, $\nu_\tau$, and $\bar\nu_e$ scattering, there are two or three possible final configurations (elastic and nonelastic subchannels). In such a case, for each subchannel, we select a point from the respective phase space independently and calculate the subchannel weights $W_i$ using the procedure described before. The overall event weight $W_{tot}$ is defined as a sum of weights $W_i$ of all available subchannels, and the event output configuration is selected to be that of $i$-th subchannel with the probability $P_i=W_i/W_{tot}$.

One can activate the new dynamics by setting the option ${\tt dyn\_lep} = 1$ in the input {\tt params.txt} file. In the output file, the neutrino-electron cross section is given in the same normalization as remaining interaction modes, i.e., `per nucleon' in the target. Samples of events with neutrino interactions on nucleons and electrons can be produced together with relative quantities determined by corresponding average cross sections.

\begin{figure}
	\centering
	\includegraphics[width=0.48\textwidth]{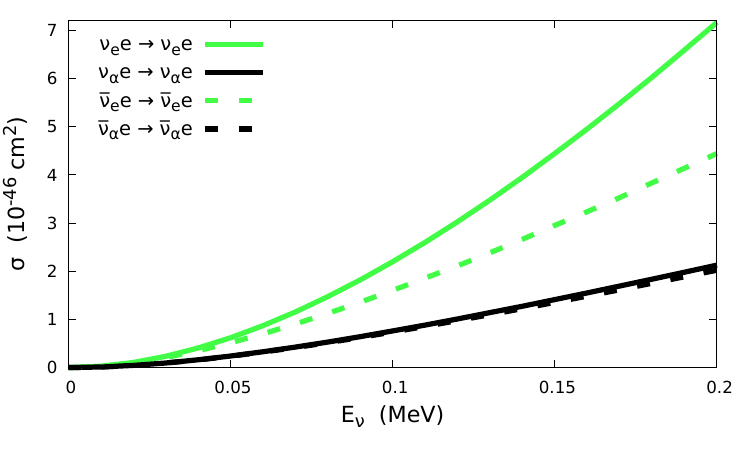}
	\includegraphics[width=0.48\textwidth]{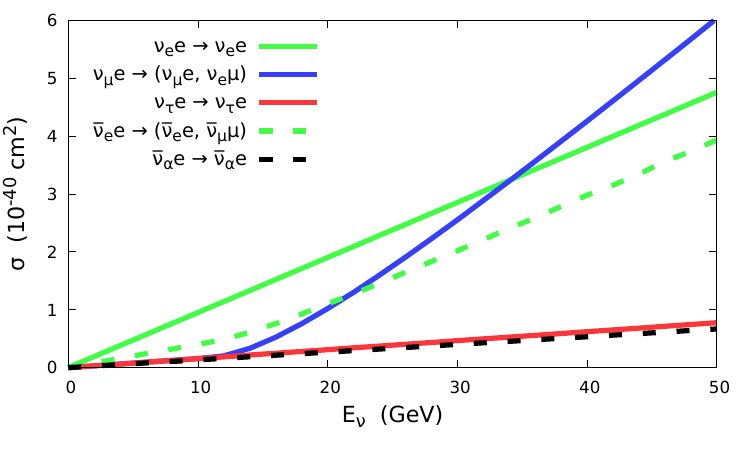}
	\includegraphics[width=0.48\textwidth]{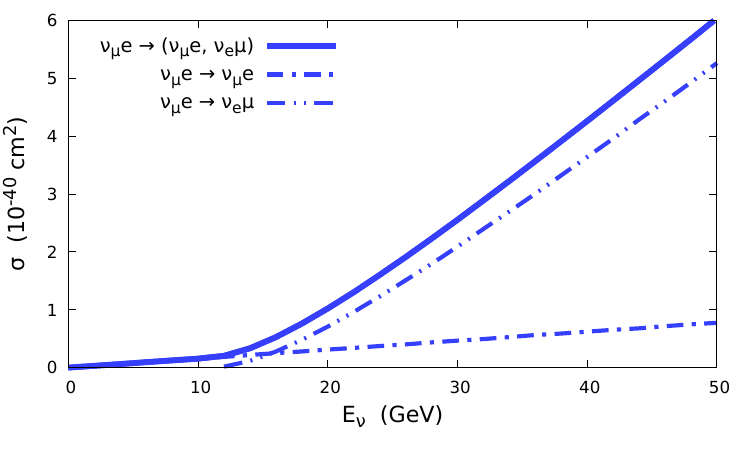}
	\includegraphics[width=0.48\textwidth]{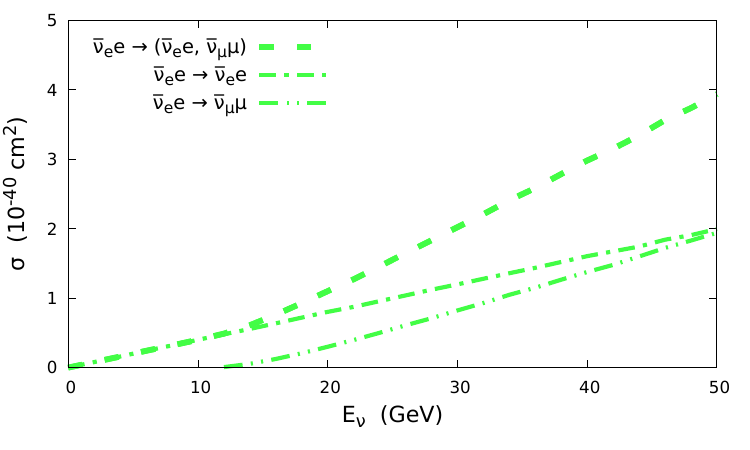}
	\vspace{-3mm}
	\caption{Cross sections for neutrino-electron reactions as a function of neutrino energy. All the results are obtained with NuWro. Two neutrino energy ranges are chosen:  (i) $E_\nu\in (0,\ 0.2)$~MeV, comparable to the electron mass (upper left), (ii) $E_\nu\in (0, 50)$~GeV (remaining three plots), allowing to see the effects around the threshold for muon production. In two lower figures, we show the breakdown of the $\nu_\mu$ (left) and $\bar\nu_e$ (right) cross sections into separate contributions from the subchannels. $\alpha$ stands for $\mu$ or $\tau$ for the processes where in a shown energy range cross sections for corresponding neutrino or antineutrino are identical. Normalization is `per electron'.
	} \label{Fig:cross-sections}
\end{figure}

In Fig.~\ref{Fig:cross-sections}, we show the total cross sections of all leptonic channels in Eq.~(\ref{eq:channels}) for the noteworthy ranges of incoming neutrino energy, including low energies where they exhibit non-linear behaviour. This is due to several reasons: the effect of electron mass (upper left entry), and the energy threshold of Eq.~(\ref{eq:thresholds}) for muon production. For the incoming neutrino energies above $\sim3$~TeV, the new subchannels with tau lepton production occur in $\nu_\tau e$ and $\bar\nu_e e$ scatterings, but they are not seen because we restricted the neutrino energy range to below $50$~GeV.

\section{Numerical analysis}
\label{sec:Applications}

We have performed an entry 
validation of our implementation by checking its ability to reproduce the analytical total cross sections given by the Eqs.~(\ref{eq:sigma1})-(\ref{eq:sigma2}), see Fig.~\ref{Fig:cross-sections}. 
In the following Subsections \ref{sec:losalamos} and \ref{sec:solar}, we compare performance of the new implementation with some published results on the accelerator and solar neutrinos. As the application, in Subsection~\ref{sec:NuWro_T2K}, we discuss the background of neutrino-electron events in the T2K $\nu_\mu\rightarrow\nu_e$ appearance measurement.

\subsection{Validation: Los Alamos accelerator data}
\label{sec:losalamos}

\begin{figure}
	\centering
	\includegraphics[width=0.8\textwidth]{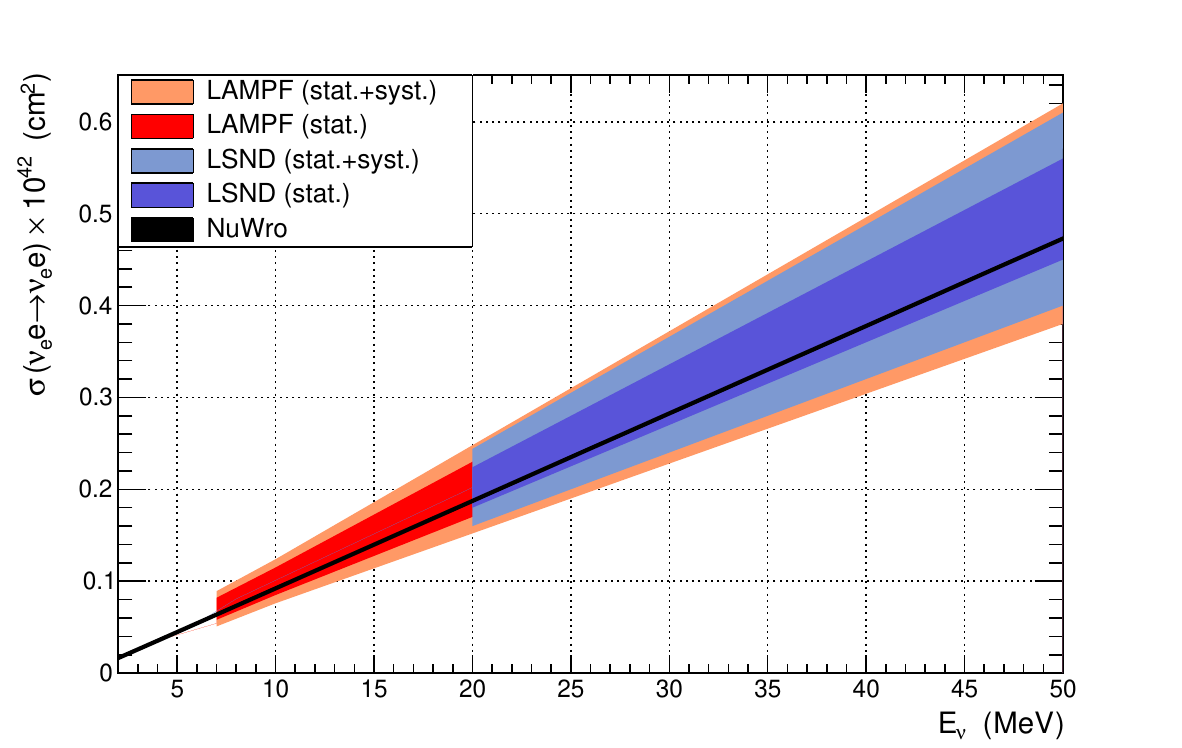}
	\vspace{-3mm}
	\caption{$\nu_e\ e\rightarrow \nu_e\ e$ scattering cross section calculated by NuWro (black line) versus the  allowed regions extracted from the data collected in LAMPF~\cite{Allen:1992qe} and LSND~\cite{Auerbach:2001wg} experiments.}
		\label{Fig:LSND}
\end{figure}

For the normalization test, we chose the measurements done by the Los Alamos National Laboratory experiments LAMPF~\cite{Allen:1992qe} and  LSND~\cite{Auerbach:2001wg}. Both of them used low energy electron neutrino beams. The reported, as energy-dependent formulas, results for $\nu_ee^-\to\nu_ee^-$ elastic scattering cross section were:
\begin{eqnarray}
    \sigma_{\rm LAMPF} &=& [10.0\pm1.5\,({\rm stat}) \pm0.9\,({\rm syst})] \times E_{\nu}({\rm MeV}) \times 10^{-45}~{\rm cm}^2, \nonumber\\
    \sigma_{\rm LSND} &=& [10.1\pm1.1\,({\rm stat}) \pm1.0\,({\rm syst})] \times E_{\nu}({\rm MeV}) \times 10^{-45}~{\rm cm}^2.
\end{eqnarray}
In Fig.~\ref{Fig:LSND}, we show experimental results as colour bands, together with the NuWro simulation results presented as a black line.  The agreement is adequate. 

\subsection{Validation: Solar neutrinos}
\label{sec:solar}

A possible application of the new NuWro interaction mode is in the studies of solar neutrinos oscillations. The detectors such as Super-Kamiokande (SK), SNO or Borexino are sensitive to the $\nu_l\,e\rightarrow \nu_l\,e$ reactions.

\begin{figure}
	\centering
	\includegraphics[width=0.8\textwidth]{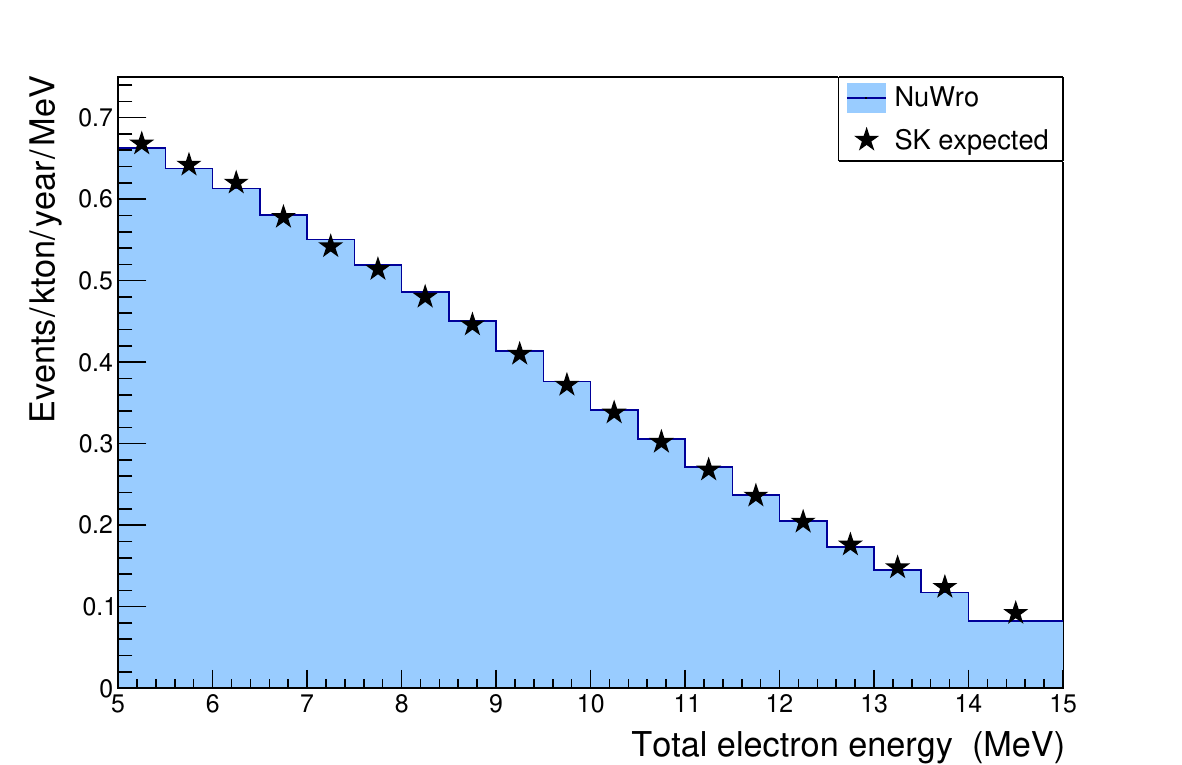}
	\vspace{-3mm}
	\caption{~The Super-Kamiokande expectation~\cite{Abe:2016nxk} marked with stars, versus NuWro distribution (blue histogram) of electron energies arising from  $\nu_e\,e\rightarrow\nu_e\,e$ scattering of the $hep$ solar neutrinos. The NuWro results are normalized to the same area, allowing for a shape-only comparison.}
	\label{Fig:hep}
\end{figure}

We have performed a comparison to the SK expectation~\cite{Abe:2016nxk} for the non-monochromatic $hep$ solar neutrinos, which come from the $^3{\rm He}+p\to{\rm ^4He}+e^++\nu_e$ reaction. The spectrum of such $hep$ neutrinos can be approximated by the formula from Ref.~\cite{Raffelt:1996wa}:
\begin{eqnarray}
    \frac{dN}{dE_\nu} = 2.33\times10^{-5}~(18.8-E_\nu)^{1.80} E_\nu^{1.92}.
\end{eqnarray}
In Fig.~\ref{Fig:hep}, the stars mark predicted event rates for the $\nu_e\,e\rightarrow\nu_e\,e$ scattering of the $hep$ neutrinos in the SK detector~\cite{Abe:2016nxk}, neglecting oscillations. The histogram shows the respectively normalized NuWro distribution for $10^7$ generated events. The agreement between the two computations is accurate.

\begin{figure}
	\centering
	\includegraphics[width=0.8\textwidth]{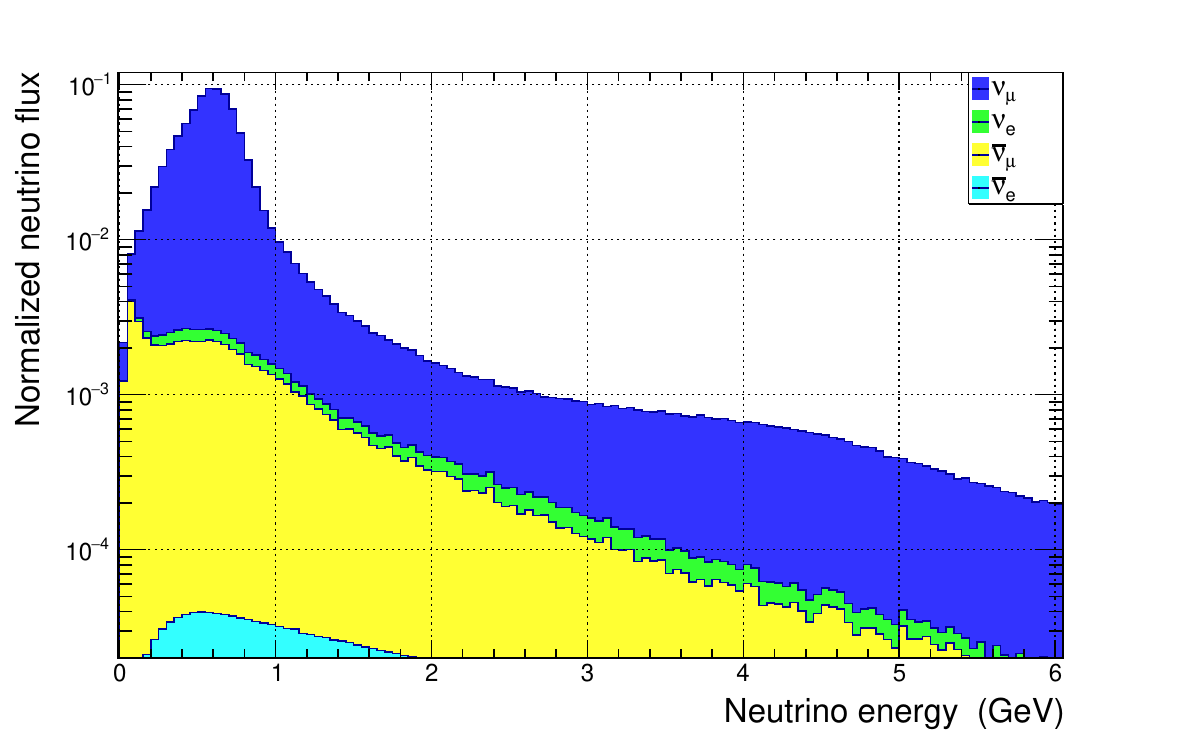}
	\includegraphics[width=0.8\textwidth]{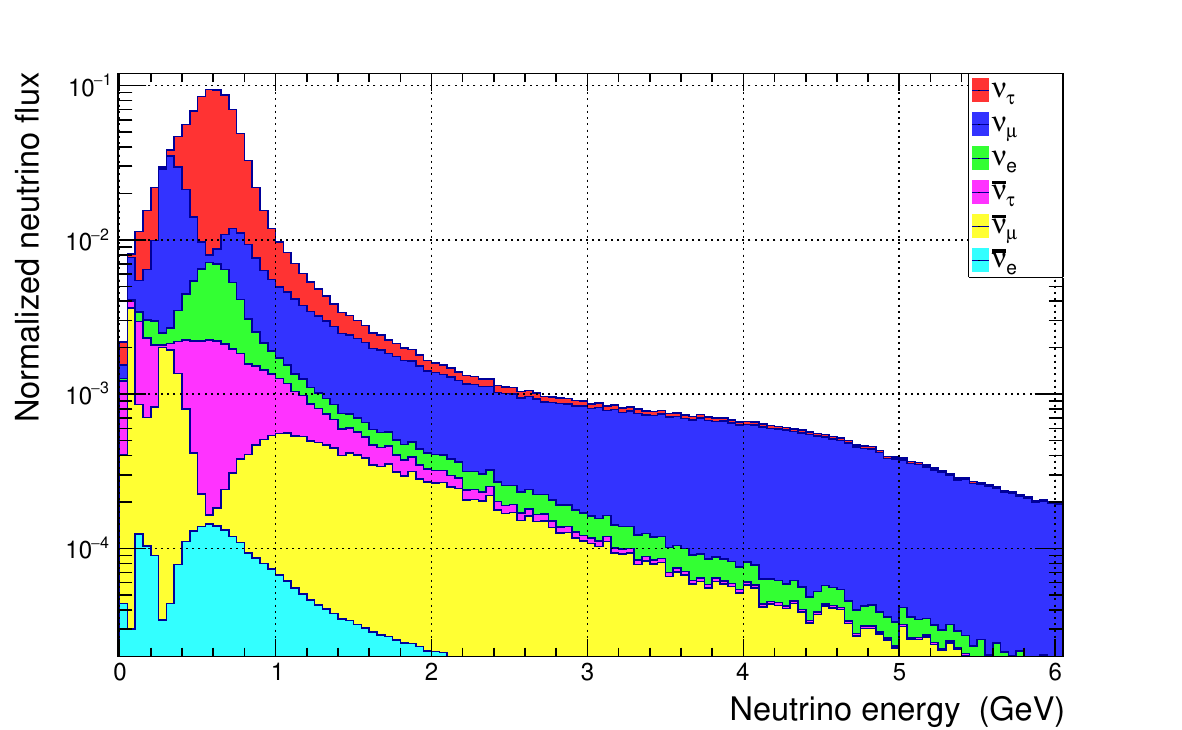}
	\vspace{-3mm}
	\caption{The unoscillated (upper) and oscillated (lower) T2K beam flux in the neutrino mode at the Super-Kamiokande detector. The oscillation parameters used in the computations are: $\sin^2\theta_{13}=2.241\times10^{-2}$, $\sin^2\theta_{23}=0.558$, and $\Delta m_{32}^2 = 2.449\times10^{-3}$~eV$^2$~\cite{Tanabashi:2018oca}.}
	\label{Fig:T2K_SK_flux}
\end{figure}

\begin{table}
		\caption{\label{Tab:results}The breakdown of electron-like events at Super-Kamiokande with the T2K neutrino flux for the same oscillation parameters as in Fig.~\ref{Fig:T2K_SK_flux}. The second row shows the composition of the flux in the energy range from 0 to 6 GeV. The fourth row shows the average cross sections for various interaction modes, and the last row shows their relative contributions to the overall number of events.}
	\begin{indented}
	\lineup
	\item[]\begin{tabular}{l|c|c|c|c|c}
			\br
			Neutrino type & \multicolumn{2}{c|}{$\nu_e$} & \ \ $\nu_\mu$ \ \ & \ \ $\nu_\tau$ \ \ & \ \ $\bar\nu$ \ \  \\
			\mr
			\ Flux contribution in \% \ & \multicolumn{2}{c|}{4.38} & 24.92 & 64.46 & 6.24 \\
			\br
			\br
			Scattering channel & \ CCQE \ & \ \ $\nu_e$-\,$e$ \ \ & \ \ $\nu_\mu$-\,$e$ \ \ & \ \ $\nu_\tau$-\,$e$ \ \ & \ \ $\bar\nu$-\,$e$ \ \ \\
			\mr
			\ $\langle\sigma\rangle\times10^{42}$ (\,cm$^2$) \ & 3220 & 4.27 & 1.07 & 0.551 & 0.833 \\
			\ Event number in \% & 99.41 & 0.13 & 0.17 & 0.25 & 0.04 \\
			\br
		\end{tabular}
	\end{indented}
\end{table}

\subsection{Application: T2K experiment}\label{sec:NuWro_T2K}

Tokai-to-Kamioka (T2K)~\cite{Abe:2011ks} is a long-baseline neutrino oscillation experiment in which the neutrinos produced in the J-PARC accelerator center in Tokai travel a distance of about 295 km, aiming to interact in the Super-Kamiokande laboratory. The neutrino flux is composed mostly of muon neutrinos or antineutrinos.

T2K performs experimental analyses aiming to measure parameters of the neutrino oscillations model~\cite{Tanabashi:2018oca}. The disappearance signal $\nu_\mu\not\rightarrow\nu_\mu$ is used to measure the $|\Delta m_{23}^2|$ and $\sin^2\theta_{23}$ oscillation parameters. The dominant oscillation mode at T2K is $\nu_\mu\rightarrow\nu_\tau$, yet the neutrino energies are too low to produce a charged lepton $\tau$ at the oscillation maximum. Instead, one can measure the subleading appearance mode $\nu_\mu\rightarrow\nu_e$, which is sensitive to the values of $\theta_{13}$ and $CP$-violating phase. Electron neutrinos are measured in the SK detector mostly via the CCQE scattering. The SK identifies electron neutrinos `electron-like' events because it can distinguish electron and muon Cherenkov rings. Hence, it is important to estimate the background coming from the $\nu\,e\rightarrow\nu\,e$ interactions.

The predicted unoscillated T2K flux in the neutrino mode, taken from Refs.~\cite{Abe:2012av,SK_beam}, is shown in the top plot of Fig.~\ref{Fig:T2K_SK_flux}, while the bottom panel presents the oscillated one. To compute the oscillated flux, we used the best fit values of neutrino oscillation parameters for the normal ordering of neutrino masses~\cite{Tanabashi:2018oca}: $\sin^2\theta_{13}=2.241\times10^{-2}$, $\sin^2\theta_{23}=0.558$, $\Delta m_{32}^2 = 2.449\times10^{-3}$~eV$^2$, and neglected the effects of the smaller parameter $\Delta m_{21}^2$, which contribution to the effective neutrino mass splitting is less than 2\%~\cite{Adamson:2013whj}, and the $CP$ violation.

\begin{figure}
	\centering
	\includegraphics[width=0.8\textwidth]{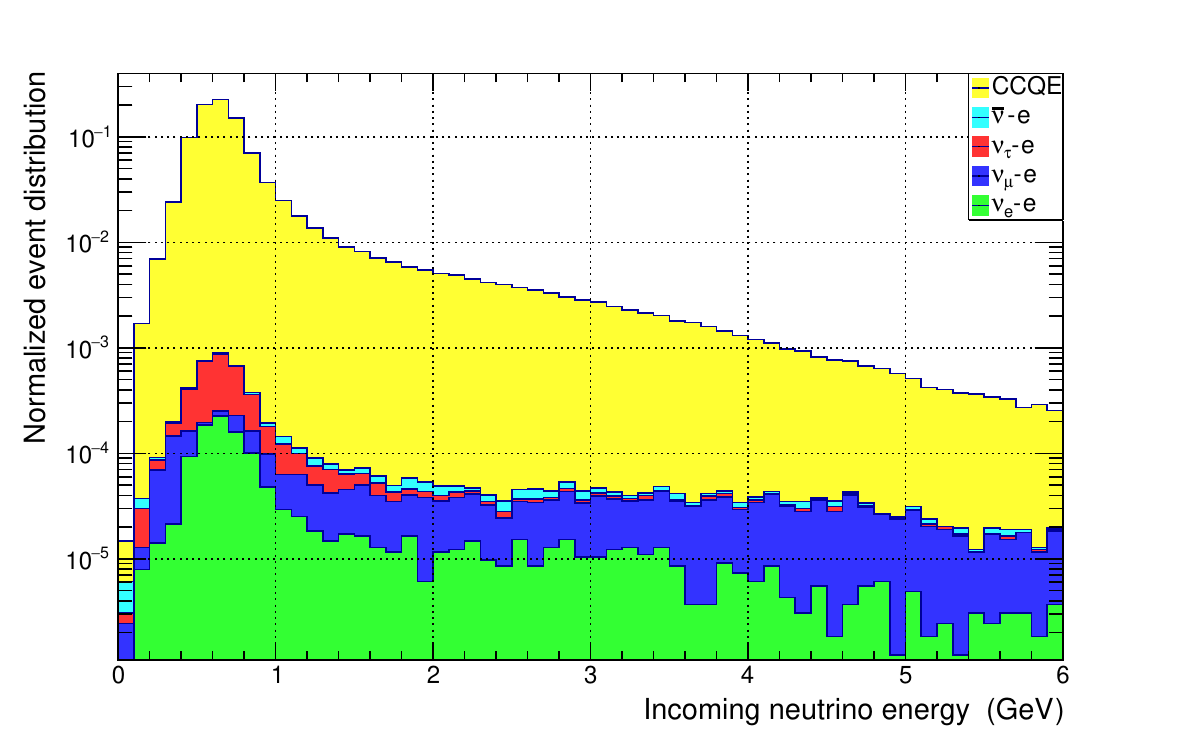}
	\vspace{-3mm}
	\caption{NuWro prediction for the T2K experiment signal of electron-like events measured in SuperKamiokande detector. For details, see Table~\ref{Tab:results}. }
	\label{Fig:T2K}
\end{figure}

In Table~\ref{Tab:results}, we collect information about the composition of the T2K neutrino beam at the SK and show the predicted fractions of electron-like events broken down according to their origin. The overall protons-on-target (POT) normalization is arbitrary, and we are concerned only about the relative fractions. We present the distribution of the electron-like events with the breakdown into interaction modes as a function of neutrino energy in Fig.~\ref{Fig:T2K}.

\begin{figure}
	\centering
	\includegraphics[width=0.8\textwidth]{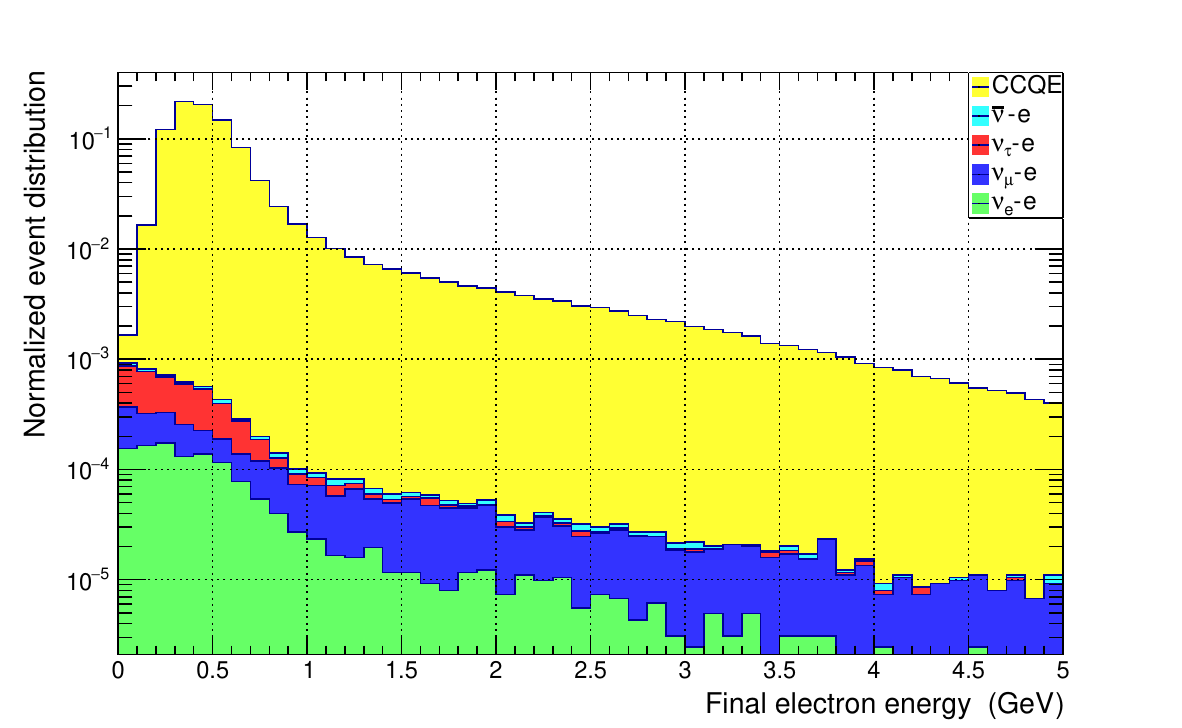}
	\vspace{-3mm}
	\caption{A distribution of electron-like events at SK with the T2K $\nu_\mu$ flux, as a function of the final electron energy. The contributions from the CCQE and purely leptonic interactions are shown separately.}
	\label{Fig:T2K_Ee}
\end{figure}

\begin{figure}
	\centering
	\includegraphics[width=0.8\textwidth]{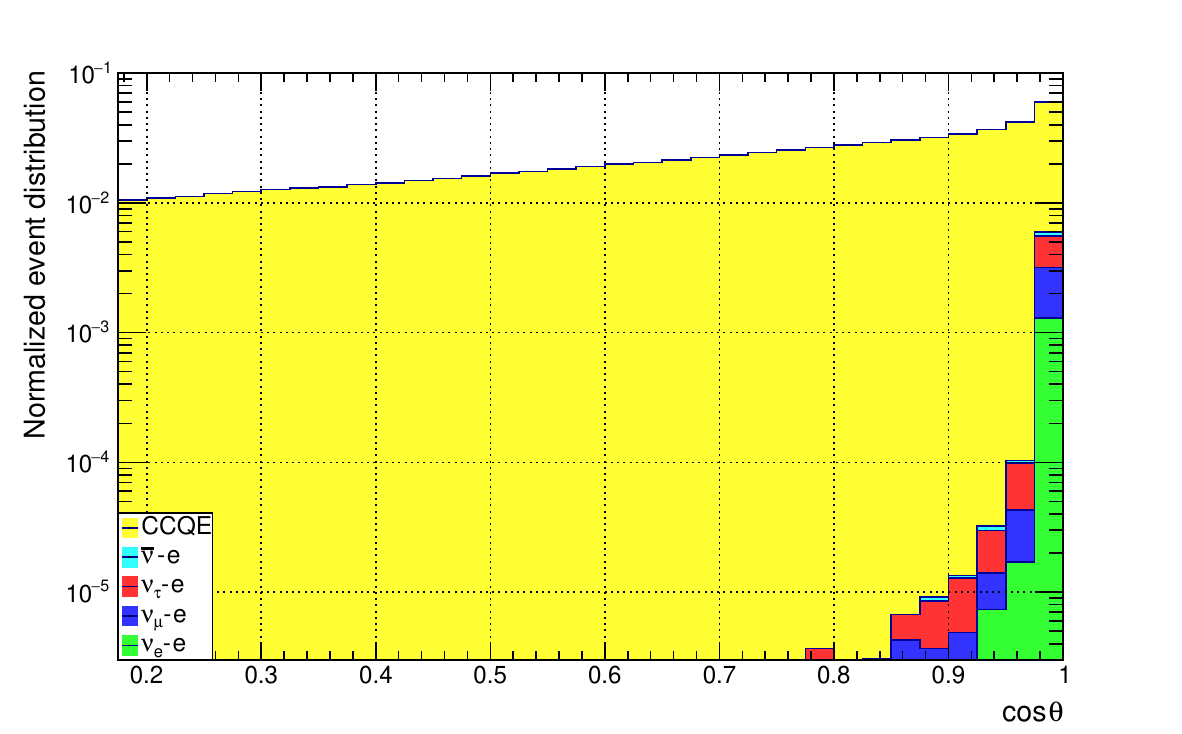}
	\vspace{-3mm}
	\caption{~The same as in Fig.~\ref{Fig:T2K_Ee} but the distribution is drawn as a function of $\cos\theta$.}
	\label{Fig:T2K_Theta}
\end{figure}

In Figs.~\ref{Fig:T2K_Ee} and \ref{Fig:T2K_Theta}, we show distributions of the signal from CCQE electron-like events and the leptonic background in the final electron kinetic energy, and the cosine of outgoing electron angle relative to the direction of the neutrino flux, respectively. From these comparisons one can deduce that the only kinematical region where the background events can contribute significantly to the overall signal is that of forward (relative to neutrino beam) moving low energy electrons. This observation can be verified with the analytic computations using the relation between the final electron angle $\theta$ and energy $E_e'$: 
\begin{eqnarray}\label{eq:cos}
    \cos\theta = \frac{E_\nu E_e'-m_e(E_\nu-E_e')-m_e^2}{E_\nu\sqrt{{E_e'}^2-m_e^2}}= \frac{\varepsilon(E_\nu+m_e)}{E_\nu\sqrt{\varepsilon^2+2m_e\varepsilon}},
\end{eqnarray}
where we introduced the electron kinetic energy $\varepsilon$: $E_e'=m_e+\varepsilon$. At T2K energies $E_\nu>\!\!>m_e$ and the relation simplifies as
\begin{eqnarray}\label{eq:cos2}
    \cos\theta \approx \frac{\varepsilon}{\sqrt{\varepsilon^2+2m_e\varepsilon}} =  
    \left( 1+\frac{2m_e}{\varepsilon} \right)^{-1/2}.
\end{eqnarray}
Hence, for $\varepsilon>\!\!>m_e$, one obtains $\cos\theta \,\approx 1-m_e/\varepsilon \approx 1$.


\begin{figure}
	\centering
	\includegraphics[width=0.8\textwidth]{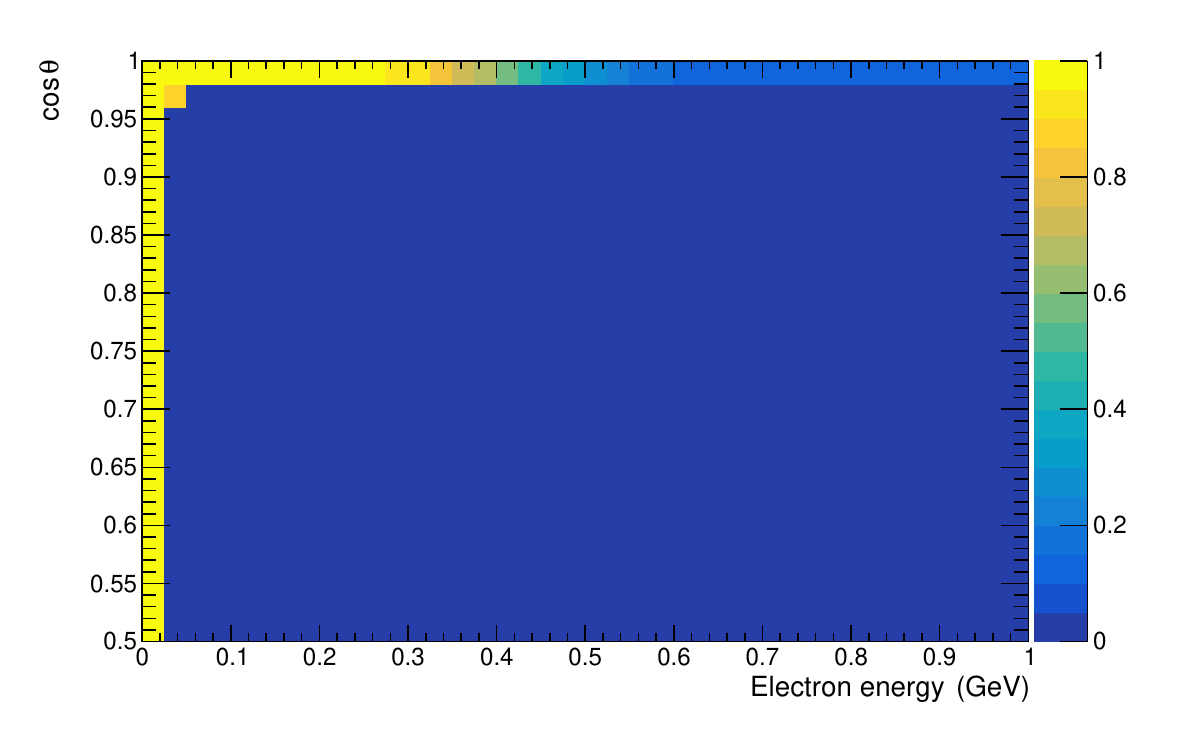}
	\vspace{-3mm}
	\caption{NuWro prediction for the two-dimensional $E_e^\prime$\,-\,$\cos\theta$ distribution of the ratio of the leptonic to total (CCQE plus leptonic) electron-like events at the Super-Kamiokande detector: $N_{\rm{LEP}}$/($N_{\rm{CCQE}}$+$N_{\rm{LEP}}$). 
	}
	\label{Fig:2D}
\end{figure}

To investigate this region of phase space in more detail, in Fig.~\ref{Fig:2D}, we show the
ratio $N_{\rm{LEP}}$/($N_{\rm{CCQE}} $+$N_{\rm{LEP}}$), where $N_{\rm{LEP}}$ is the number of neutrino-electron events and  $N_{\rm{CCQE}} $ is the number of CCQE events, calculated in 2D $(E_e^\prime ,\ \cos\theta)$ bins. One can see that the neutrino-electron and the CCQE events are mostly separated. The regions which are almost entirely populated by neutrino-electron events, i.e., those in the most forward bin and with low values of the final state electron energy $E_e^\prime\leq 300$~MeV account for $\sim 40\%$ of the total number of $N_{\rm{LEP}}$ events. With a sufficient detector resolution in terms of electron energy and angle, one could reject this fraction of neutrino-electron events. Assuming a total statistics of the order of $10^{22}$~POT by the end of T2K commissioning, we expect to find a few neutrino-electron events. In the future Hyper-Kamiokande oscillation experiment with the statistics higher by a factor of 20, this number will grow, making such considerations reasonable. We showed that almost half of this background to electron-like events populate a distinct region of the phase space and can be subtracted. The complete study of the background events in SK measurement of the appearance oscillation signal goes far beyond the scope of this paper. Background events also originate from the NC $\pi^0$ and single-photon production. The former plays a crucial role in the experimental analyses, and there have been many exhaustive studies of methods to distinguish $e$ and $\pi^0$ signatures at SK~\cite{Abe:2013hdq}. Here, we emphasize a source of the background that can result in a similar final state already at the interaction vertex level and cannot be tackled otherwise. For the latter, the Ref.~\cite{Wang:2015ivq} yielded a calculated amount of about $1.5\%$ of the overall oscillation signal coming from this mechanism. Photons arising from the NC$1\gamma$ reaction populate a large region in the $(E_\gamma, \cos\theta_\gamma)$ plane and contrary to the $\nu\ e$ scattering events, cannot be kinematically separated.

Finally, we notice that due to radiative corrections, the fraction of neutrino-electron events in the low electron energy bins becomes slightly higher. As explained in Sec.~\ref{sec:Analytics}, electrons tend to loose energy due to emission of bremsstrahlung photons~\cite{Sarantakos:1982bp}.

\section{Conclusions}\label{sec:Conclusion}

In this paper, we describe the implementation of neutrino-electron reactions in the NuWro Monte Carlo neutrino event generator. Using this new functionality, we compared the NuWro outcome with a selected sample of experimental data and theoretical computations. The most elucidative example is that of electron-like events detected in the Super-Kamiokande detector in the T2K experiment. Using NuWro results, we argue that in the future Hyper-Kamiokande oscillation experiment, such interactions should be included in the analysis.

\section*{Acknowledgements} 

We thank Krzysztof Graczyk for useful discussions. The authors were supported by NCN Opus Grant 2016/21/B/ST2/01092. KN was partially supported by the Special Research Fund, Ghent University, and together with JS, by the Polish Ministry of Science and Higher Education, Grant No. DIR/WK/2017/05.

\section*{References}
\bibliography{bibliography_LEP_1}

\providecommand{\newblock}{}
\begin{thebibliography}{10}
\expandafter\ifx\csname url\endcsname\relax
  \def\url#1{{\tt #1}}\fi
\expandafter\ifx\csname urlprefix\endcsname\relax\def\urlprefix{URL }\fi
\providecommand{\eprint}[2][]{\url{#2}}

\bibitem{Abe:2019vii}
Abe K {\em et~al.\/} (T2K) 2020 {\em Nature\/} {\bf 580} 339--344
  (\textit{Preprint} \eprint{arXiv:1910.03887})

\bibitem{Abi:2020evt}
Abi B {\em et~al.\/} (DUNE) 2020  (\textit{Preprint} \eprint{arXiv:2002.03005})

\bibitem{Abe:2015zbg}
Abe K {\em et~al.\/} (Hyper-Kamiokande Proto-) 2015 {\em PTEP\/} {\bf 2015}
  053C02 (\textit{Preprint} \eprint{arXiv:1502.05199})

\bibitem{Fukuda:2001nj}
Fukuda S {\em et~al.\/} (Super-Kamiokande) 2001 {\em Phys. Rev. Lett.\/} {\bf
  86} 5651--5655 (\textit{Preprint} \eprint{arXiv:hep-ex/0103032})

\bibitem{BOGER2000172-short}
Boger J {\em et~al.\/} (SNO) 2000 {\em Nucl. Instrum. Meth. A\/} {\bf 449} 172
  -- 207

\bibitem{Alimonti:2008gc}
Alimonti G {\em et~al.\/} (Borexino) 2009 {\em Nucl. Instrum. Meth. A\/} {\bf
  600} 568--593 (\textit{Preprint} \eprint{arXiv:0806.2400})

\bibitem{Bionta:1987qt}
Bionta R {\em et~al.\/} 1987 {\em Phys. Rev. Lett.\/} {\bf 58} 1494

\bibitem{Tanabashi:2018oca}
Tanabashi M {\em et~al.\/} (Particle Data Group) 2018 {\em Phys. Rev. D\/} {\bf
  98} 030001

\bibitem{Hayato:2009zz}
Hayato Y 2009 {\em Acta Phys. Polon. B\/} {\bf 40} 2477--2489

\bibitem{Andreopoulos:2009rq}
Andreopoulos C {\em et~al.\/} 2010 {\em Nucl. Instrum. Meth. A\/} {\bf 614}
  87--104 (\textit{Preprint} \eprint{arXiv:0905.2517})

\bibitem{NuWro}
 {\em NuWro repository \url{https://github.com/NuWro/nuwro}\/}

\bibitem{GiBUUrev}
Buss O, Gaitanos T, Gallmeister K, van Hees H, Kaskulov M, Lalakulich O,
  Larionov A, Leitner T, Weil J and Mosel U 2012 {\em Phys. Rept.\/} {\bf 512}
  1--124 (\textit{Preprint} \eprint{arXiv:1106.1344})

\bibitem{Alvarez-Ruso:2017oui}
Alvarez-Ruso L {\em et~al.\/} (NuSTEC) 2018 {\em Prog. Part. Nucl. Phys.\/}
  {\bf 100} 1--68 (\textit{Preprint} \eprint{arXiv:1706.03621})

\bibitem{Grichine:2019vcc}
Grichine V~M 2019 {\em Nucl. Instrum. Meth. A\/} {\bf 942} 162403

\bibitem{Park:2015eqa}
Park J {\em et~al.\/} (MINERvA) 2016 {\em Phys. Rev. D\/} {\bf 93} 112007
  (\textit{Preprint} \eprint{arXiv:1512.07699})

\bibitem{Valencia:2019mkf}
Valencia E {\em et~al.\/} (MINERvA) 2019 {\em Phys. Rev. D\/} {\bf 100} 092001
  (\textit{Preprint} \eprint{arXiv:1906.00111})

\bibitem{Abe:2011ks}
Abe K {\em et~al.\/} (T2K) 2011 {\em Nucl. Instrum. Meth. A\/} {\bf 659}
  106--135 (\textit{Preprint} \eprint{arXiv:1106.1238})

\bibitem{Tomalak:2019ibg}
Tomalak O and Hill R~J 2020 {\em Phys. Rev. D\/} {\bf 101} 033006
  (\textit{Preprint} \eprint{arXiv:1907.03379})

\bibitem{Fermi:1934hr}
Fermi E 1934 {\em Z. Phys.\/} {\bf 88} 161--177

\bibitem{Healey:2013vka}
Healey K~J, Petrov A~A and Zhuridov D 2013 {\em Phys. Rev. D\/} {\bf 87} 117301
  [Erratum: Phys.Rev.D 89, 059904 (2014)] (\textit{Preprint}
  \eprint{arXiv:1305.0584})

\bibitem{Glashow:1960zz}
Glashow S~L 1960 {\em Phys. Rev.\/} {\bf 118} 316--317

\bibitem{Garcia:2020jwr}
Garcia A, Gauld R, Heijboer A and Rojo J 2020  (\textit{Preprint}
  \eprint{arXiv:2004.04756})

\bibitem{tHooft:1971ucy}
't~Hooft G 1971 {\em Phys. Lett. B\/} {\bf 37} 195--196

\bibitem{Sarantakos:1982bp}
Sarantakos S, Sirlin A and Marciano W 1983 {\em Nucl. Phys. B\/} {\bf 217}
  84--116

\bibitem{Lee:1964jq}
Lee T and Sirlin A 1964 {\em Rev. Mod. Phys.\/} {\bf 36} 666--669

\bibitem{Marciano:2003eq}
Marciano W~J and Parsa Z 2003 {\em J. Phys. G\/} {\bf 29} 2629--2645
  (\textit{Preprint} \eprint{arXiv:hep-ph/0403168})

\bibitem{NuWroFSI}
Golan T, Juszczak C and Sobczyk J~T 2012 {\em Phys. Rev. C\/} {\bf 86}(1)
  015505

\bibitem{Niewczas:2019fro}
Niewczas K and Sobczyk J~T 2019 {\em Phys. Rev. C\/} {\bf 100} 015505
  (\textit{Preprint} \eprint{arXiv:1902.05618})

\bibitem{Allen:1992qe}
Allen R {\em et~al.\/} 1993 {\em Phys. Rev. D\/} {\bf 47} 11--28

\bibitem{Auerbach:2001wg}
Auerbach L {\em et~al.\/} (LSND) 2001 {\em Phys. Rev. D\/} {\bf 63} 112001
  (\textit{Preprint} \eprint{arXiv:hep-ex/0101039})

\bibitem{Abe:2016nxk}
Abe K {\em et~al.\/} (Super-Kamiokande) 2016 {\em Phys. Rev. D\/} {\bf 94}
  052010 (\textit{Preprint} \eprint{arXiv:1606.07538})

\bibitem{Raffelt:1996wa}
Raffelt G 1996 {\em {Stars as laboratories for fundamental physics}\/} ISBN
  978-0-226-70272-8

\bibitem{Abe:2012av}
Abe K {\em et~al.\/} (T2K) 2013 {\em Phys. Rev. D\/} {\bf 87} 012001 [Addendum:
  Phys.Rev.D 87, 019902 (2013)] (\textit{Preprint} \eprint{arXiv:1211.0469})

\bibitem{SK_beam}
 {\em T2K flux at SK detector can be found at:
  \url{https://t2k-experiment.org/result\_category/flux}\/}

\bibitem{Adamson:2013whj}
Adamson P {\em et~al.\/} (MINOS) 2013 {\em Phys. Rev. Lett.\/} {\bf 110} 251801
  (\textit{Preprint} \eprint{arXiv:1304.6335})

\bibitem{Abe:2013hdq}
Abe K {\em et~al.\/} (T2K) 2014 {\em Phys. Rev. Lett.\/} {\bf 112} 061802
  (\textit{Preprint} \eprint{arXiv:1311.4750})

\bibitem{Wang:2015ivq}
Wang E, Alvarez-Ruso L, Hayato Y, Mahn K and Nieves J 2015 {\em Phys. Rev. D\/}
  {\bf 92} 053005 (\textit{Preprint} \eprint{arXiv:1507.02446})

\end{thebibliography}

\end{document}